\documentclass[a4paper,fleqn,usenatbib]{mnras}

\usepackage[T1]{fontenc}
\usepackage{ae,aecompl}

\usepackage{graphicx}	
\usepackage{amsmath}	
\usepackage{amssymb}	

\title[Origins of hot minerals]{Dust processing in protoplanetary envelopes as the origin of hot minerals in comets}

\author[M. Ali-Dib ]{
Mohamad Ali-Dib$^{1}$\thanks{E-mail: malidib@nyu.edu}\\
$^{1}$Center for Astro, Particle and Planetary Physics (CAP3), New York University Abu Dhabi, UAE
}

\date{Accepted 2022-12-30}

\pubyear{2023}

\begin{document}
\label{firstpage}
\pagerange{\pageref{firstpage}--\pageref{lastpage}}
\maketitle

\begin{abstract}
Crystalline silicates are found in a large number of comets. These pose a long-standing conundrum for solar system formation models as they can only be created in the inner hot disk at temperatures higher than 800 K, and there is no obvious mechanism to transport them out into the comets formation region. Here we propose that these particles could have formed inside the hydrostatic envelopes surrounding young protoplanets still embedded in the protoplanetary disk. Using a simplified 1D model we investigate the thermal structure of these envelopes, and find that for core masses ranging from 0.08 to 1.5 M$_\oplus$, located anywhere between 1 and 30 AU, the temperature and pressure at the base of the envelopes are high enough to quickly vaporize silicate particles of various sizes. Moreover, if the grain abundance is atleast solar, these envelopes become fully convective, allowing for dust ejection across the Bondi radius back into the disk. Amorphous silicates are hence thermally processed into crystalline particles in these envelopes, and then transported back to disk through convective diffusion to be finally incorporated into the cometary building blocks. 

\end{abstract}



\section{Introduction}
High temperature minerals are ubiquitous in the cold outer solar system small bodies. One of the earliest remote sensing detections was for crystalline silicates (CSs) such as olivines and pyroxenes in the grains of comets 1P/Halley, D/1993 F2 (Shoemaker-Levy), C/1987 P1 (Bradfield) \citep{hanner1}, and C/1993 A1 (Mueller) \citep{hanner2}. Subsequent detections were made in comets C/1995 O1 (Hale-Bopp) \citep{hayward,crovisier1997,wooden}, 103P/Hartley \citep{crovisier2000}, and more recently 17P/Holmes \citep{shinnaka}. On the other hand Calcium-Aluminum inclusions (CAIs) that form at even higher temperatures were found in the dust collected by \textit{Stardust} in comet 81P/Wild \citep{brownlee}. We refer the reader to the observational review of \cite{mumma} for more informations. 

The presence of CSs have been a primary challenge to solar system formation models for decades, as the thermal conditions in the outer protoplanetary disk are not conducive to their formation locally. {CSs can form starting from amorphous silicates through either direct vaporization followed by re-condensation at temperatures higher than $\sim$1800 K, or thermal annealing for T$>800$ K. Annealing is a physical process where sufficiently energetic molecules of a solid slowly regroup into a crystal lattice. This mechanism is not instantaneous and necessitates high temperature exposure for a period of few weeks followed by slow cooling \citep{gail1998,gail2001}. In typical protoplanetary disks, direct condensation can be active only inside $\sim$ 0.1 AU, and annealing is inefficient outside $\sim$ 1.5 AU.}

Moreover, there is no acceptable mechanism to transport these particles from the inner disk to the comets formation region.
Earlier transport models relied on turbulent diffusion \citep{morvan}, but recent ALMA observations suggest that protoplanetary disks are laminar \citep{Flaherty2015,Flaherty2018}, and thus this mechanism is unlikely to be efficient. Large scale outward advection in the disk's midplane has also been proposed \citep{hughes}, but 3D MHD simulations ruled out the presence of such advection \citep{fromang}. Another possible transport mechanism is photophoresis \citep{mousis}, but this necessitates a relatively large (1-2 AU) central hole in the disk. Finally, \cite{alidib2015} proposed that FU-Ori outbursts might form these particles in situ, but this depends on the outburst trigger radius being large enough, which is uncertain.

Here we show how high temperature minerals form naturally, and in-situ, in the envelopes surrounding low mass proto-planets embedded in the disk. We present models showing that the temperatures and pressures at the base of these envelopes easily reach conditions that allow for the formation of crystalline silicates through direct vaporization and re-condensation. Primordial amorphous silicates are thus accreted and then thermally processed in these envelopes, before finally getting ejected back to the disk as crystalline particles via convective diffusion. {We emphasize that this work concerns the formation of generic CSs such as olivines and pyroxenes, and not necessarily chondrules and CAIs due to their additional formation-time constrains that are outside the scope of this work. We present our model in section 2, results in section 3, and conclude in section 4. }

\section{Model}

We model the atmosphere using the standard atmospheric structure equations. The equation of hydrostatic equilibrium is given by:

\begin{equation}
\frac{dP}{dr} = -\frac{GM}{r^2}\rho(r)
\end{equation}
{and we define the temperature gradient equation starting from the standard assumption that heat can be transported using either radiation (if the local envelope is convectively stable) or adiabatic convection (if unstable). It is hence written as: } 
\begin{equation}
\frac{dT}{dr} = \nabla \frac{T}{P} \frac{dP}{dr}
\end{equation}
{where $\nabla$ is defined, starting from the Schwarzschild convective stability criterion ($\nabla_{\mathrm{rad}}<\nabla_{\mathrm{ad}}$), to be $\nabla = \text{min}(\nabla_{\mathrm{ad}}, \nabla_{\mathrm{rad}})$. Here $\nabla_{\mathrm{ad}}$ is the adiabatic gradient}:

\begin{equation}
\nabla_{\mathrm{ad}} \equiv\left(\frac{d \ln T}{d \ln P}\right)_{\mathrm{ad}} = \frac{\gamma - 1}{\gamma}
\end{equation}
where the adiabatic constant $\gamma = 1.5$. $\nabla_{\mathrm{rad}}$ is the radiative gradient:

\begin{equation}
\nabla_{\mathrm{rad}} \equiv \frac{3 \kappa P}{64 \pi G M\sigma T^{4}} L
\end{equation}
where $L$ is the envelope's luminosity generated by accretion at a rate $\dot{M}_{acc}$ :

\begin{equation}
L=\frac{G M \dot{M}_{acc}}{R_c}
\end{equation}
 {$R_c$ is the core radius,} and $\kappa$ is the opacity that we define following \cite{ormel} as:
\begin{equation}
\kappa=\kappa_{\mathrm{gas}}+\kappa_{\mathrm{gr}}
\end{equation}
with:
\begin{equation}
\label{grainopeq}
\kappa_{\mathrm{gr}}=\kappa_{\mathrm{geom}} Q_{e} = \frac{3Z_{gr}}{4\rho_s a}\times min( \frac{0.6\pi a}{\lambda_{max}} , 2)
\end{equation}
where $Z_{gr}$ is the grains abundance, $a_s$ their size, and $\rho_s$ their internal density. we use $\rho_s = \ 3 \ g/cm^3$ for both the core and the dust particles.

The equilibrium dust size in the envelope $a_s$ is set by two competing processes: grain growth through coagulation \citep{ormel} and grain collisional destruction \citep{alidibthompson}. The relative relevance of these two processes is decided mainly by whether the collisional speeds reach the silicate fragmentation threshold ($V_f \sim$ 100 cm/s, \cite{blum}). The collisional speed is approximated here as the largest among the dust's convective velocity $V_{\mathrm{con,d}}$ (eq. \ref{vcondust}) and the dust's radial drift velocity:
\begin{equation}
V_{\mathrm{drift,d}} = \tau_{\mathrm{stop}} \frac{GM}{r^2}
\end{equation}
where $\tau_{\mathrm{stop}}$ is the stopping time.


As discussed in \citep{alidibthompson}, collisions in these envelopes are likely to be destructive. This leads to a small characteristic dust size, increasing the opacity (thus growing the convective zone), and decreasing the vaporization timescale. Here we only select models where max($V_{\mathrm{con,d}}$,$V_{\mathrm{drift,d}}$) is higher than 100 cm/s everywhere in the disk.  

The convective fragmentation dust size is hence calculated following \citep{alidibthompson} as:
\begin{equation}
\label{aconveq}
a_{s,conv} = \frac{4\pi V_f^2r^3\rho_g^2c_g}{L\rho_s}
\end{equation}
where $\rho_g$ and $c_g$ are the gas' density and sound speed. The drift fragmentation dust size is given by:
\begin{equation}
a_{s,drift} = \frac{V_f r^2 \rho_g c_g}{GM\rho_s}
\end{equation}
with finally $a_{s}$=min($a_{s,conv}$,$a_{s,drift}$).

Note that $a_{s,conv}$ is defined everywhere in the envelope, since, as discussed below, we also only select fully convective envelopes. 
For this approach to be applicable, the particles need to reach the local fragmentation threshold at every point in the envelope. Therefore, for self-consistency, we only keep models where the mean free time for collisions is shorter than the convective timescale. In the convective fragmentation regime this can be written as:
\begin{equation}
\label{cond1}
\frac{a_s}{a_s+4 \ell_g / 9}<9 Z_{gr}^2 \frac{\rho_g}{\rho_s} \mathcal{M}_{\operatorname{con}} \frac{r}{\ell_g}
\end{equation}
where $\mathcal{M}_{\operatorname{con}}$ is the convective Mach number, and $\ell_g$ the mean free path of the gas. In the drift regime this is replaced by :
\begin{equation}
\label{cond2}
\frac{3}{4 Z_{gr}} \frac{c_g^2 r}{G M} \frac{\mathcal{M}_{\mathrm{con}}}{1+9 a_s/ 4 \ell_g}<1
\end{equation}

The gas opacity is given by:
\begin{equation}
\kappa_{\mathrm{gas}}=10^{-8} \rho^{2 / 3}_{g} T^{3}
\end{equation}
Finally we close the system with the ideal gas equation of state $P=\rho_g k_B T/\mu$. We solve these equations by integrating inwards from the outer boundary at R$_{\rm out}$, the minimum of the Bondi and Hill radii, to the core. We assumed the disk is radiative and calculate its temperature and density following \cite{alidibcumming}:
\begin{equation}
T_d = 373 \ r_{au}^{-9/10} K \\ \ \ \mathrm{and} \ \  
\rho_d = 1.7\times 10^{-10} \ r_{au}^{-33/20} g/cm^3
\end{equation}

Once we have the envelope's thermal structure, we can calculate additional quantities needed for the subsequent analysis. 
We calculate the silicate particles vaporization rate as :
\begin{equation}
\frac{1}{a_s}\frac{d a_s}{d t}=-\left(\frac{\mu_{\mathrm{Sil}}}{2 \pi k T}\right)^{1 / 2} \frac{P_{\mathrm{Sil}}^{\mathrm{sat}}}{\rho_s a_s}
\end{equation}
with \citep{Krieger}: 
\begin{equation}
P_{\mathrm{sil}}^{\mathrm{sat}}(T)=3.2 \times 10^{14} e^{-\left(6 \times 10^4 \mathrm{~K}\right) / T}
\end{equation}
and hence the silicate grains vaporization timescale is given by :
\begin{equation}
 \tau_{vap,sil} = \bigg(\frac{1}{a_s}\frac{d a_s}{d t}\bigg)^{-1}
\end{equation}

We define the gas and dust convective velocities respectively as:
\begin{equation}
V_{\mathrm{con,g}}= \bigg(\frac{L}{4\pi r^2 \rho_g}\bigg)^{1/3}
\end{equation}
where we assumed that in the convective zone the energy is entirely transported through adiabatic convection, and
\begin{equation}
\label{vcondust}
V_{\mathrm{con,d}} \sim V_{\mathrm{con}}\left(V_{\mathrm{con}} \tau_{\mathrm{stop}} / r\right)^{1 / 2}
\end{equation}

We finally calculate the  dust's convective mixing timescale as:
\begin{equation}
 \tau_{mix,d} = H/V_{\mathrm{con,d}}
\end{equation}

\section{Results}
We start by exploring parameter space in order to find the values that allow for the creation of CSs in proto-envelopes. We explore core masses ranging from Pluto's mass (0.002 M$_\oplus$) to a hypothetical giant planet's core (10 M$_\oplus$), placed between 1 and 30 AU where ambient temperatures are too low to create CSs in the disk. The grains abundance $Z_{gr}$ ranges from subsolar (10$^{-3}$) to supersolar (1.0).

\begin{figure*}
\begin{centering}
	\includegraphics[scale=0.300]{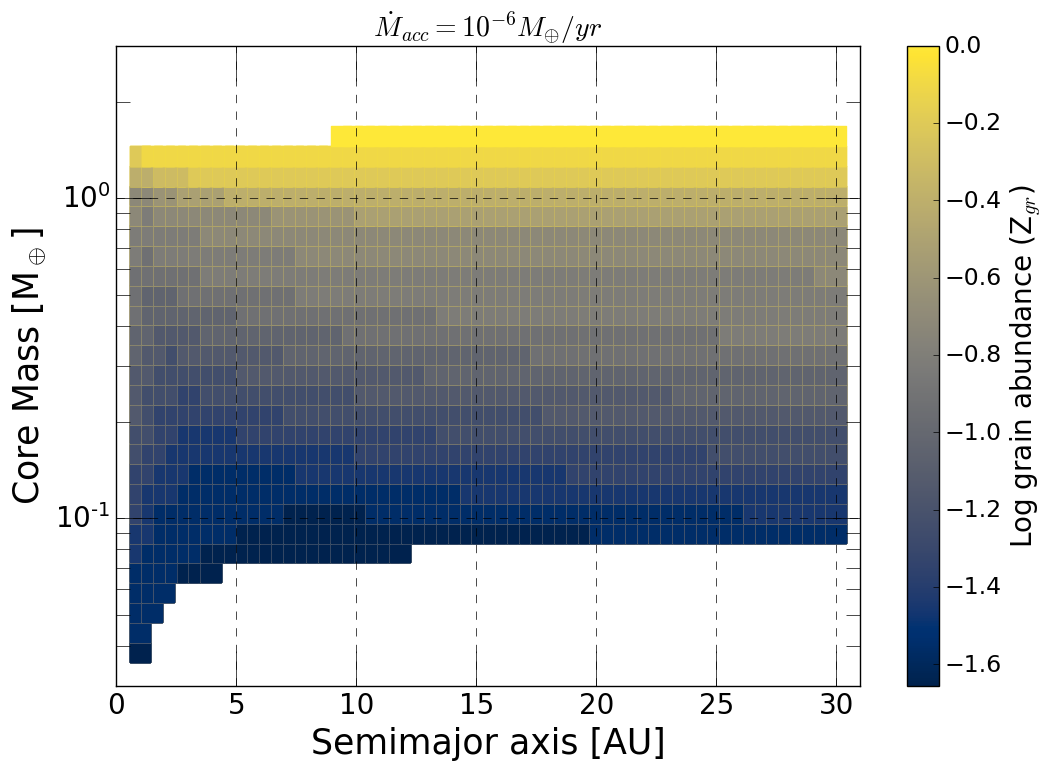}
	\includegraphics[scale=0.300]{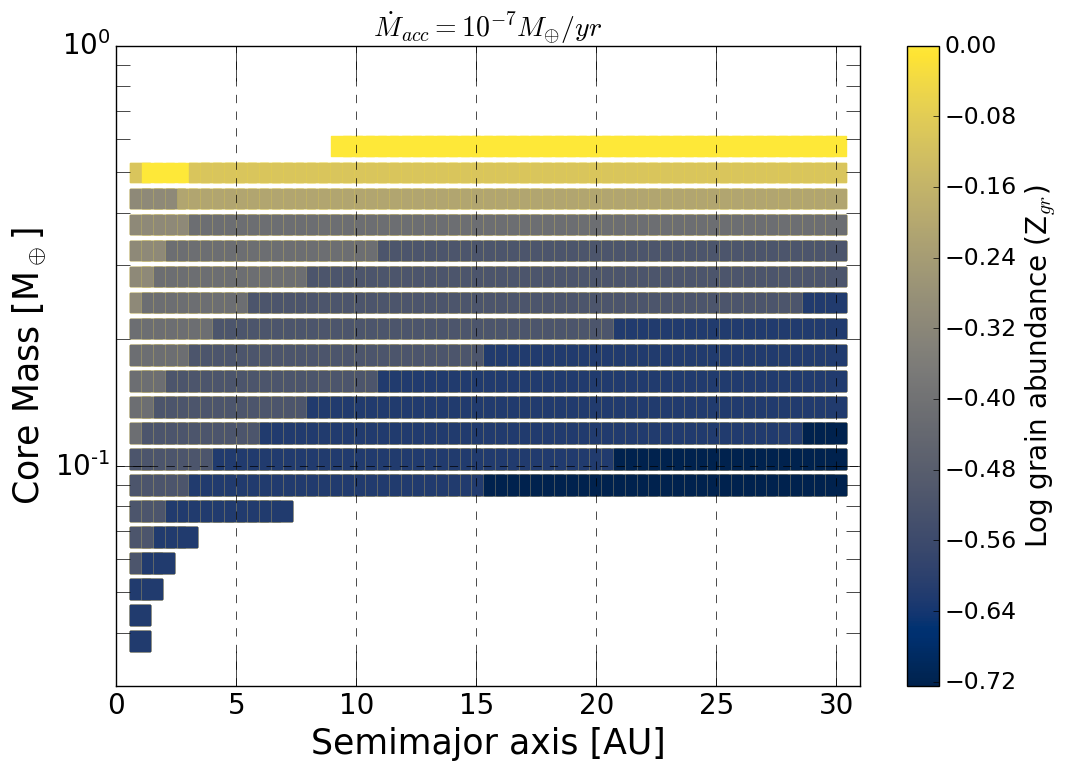}
   \caption{The semimajor axis, core mass, and envelope grain abundance for all the cases that satisfy our conditions to form and eject crystal silicates as enumerated in section 3. Left: $\dot{M}_{acc} = 10^{-6} M_{\oplus}/yr$  . Right: $\dot{M}_{acc} = 10^{-7} M_{\oplus}/yr$. {Note the different color scales for the two panels. }}
    \label{fig:main}
    \end{centering}
\end{figure*}

Our results are summarized in Fig. \ref{fig:main}. In this plot we show only the areas of parameter space leading to envelopes conducive to the creation of CSs and that are self-consistent to our model assumptions. This is defined by these conditions:
\begin{enumerate}
\item $\tau_{vap,sil}$ is less than $\tau_{mix,d}$ at the base of the envelope. This simply constrain the envelopes to those where  solid silicates at their base can get vaporized faster than they are transported back into the upper cooler zones.
\item The envelope is fully convective. This ensures that the newly created CSs can be convectively diffused all the way back into the disk. This condition is inspired by the results of \cite{alidibthompson} who considered a similar setup with a 0.3 M$_\oplus$ core embedded in the disk, and showed that, for typical accretion rates, pebble fragmentation and dust loading increases the opacity and push the convective zone out till it reaches the Bondi radius. Dust particles in these steady-state envelopes are then diffusively ejected back to the disk. Our results rely on this mechanism to transport the newly created CSs from the  hot inner envelope back to the disk to be incorporated in proto-comets. 
\item The collisional velocity is higher than 100 cm/s throughout the envelope, and conditions \ref{cond1} and \ref{cond2} are satisfied. This ensures that our dust size prescription is self-consistent.
\end{enumerate}

\subsection{Core mass}
\begin{figure*}
\begin{centering}
	\includegraphics[scale=0.250]{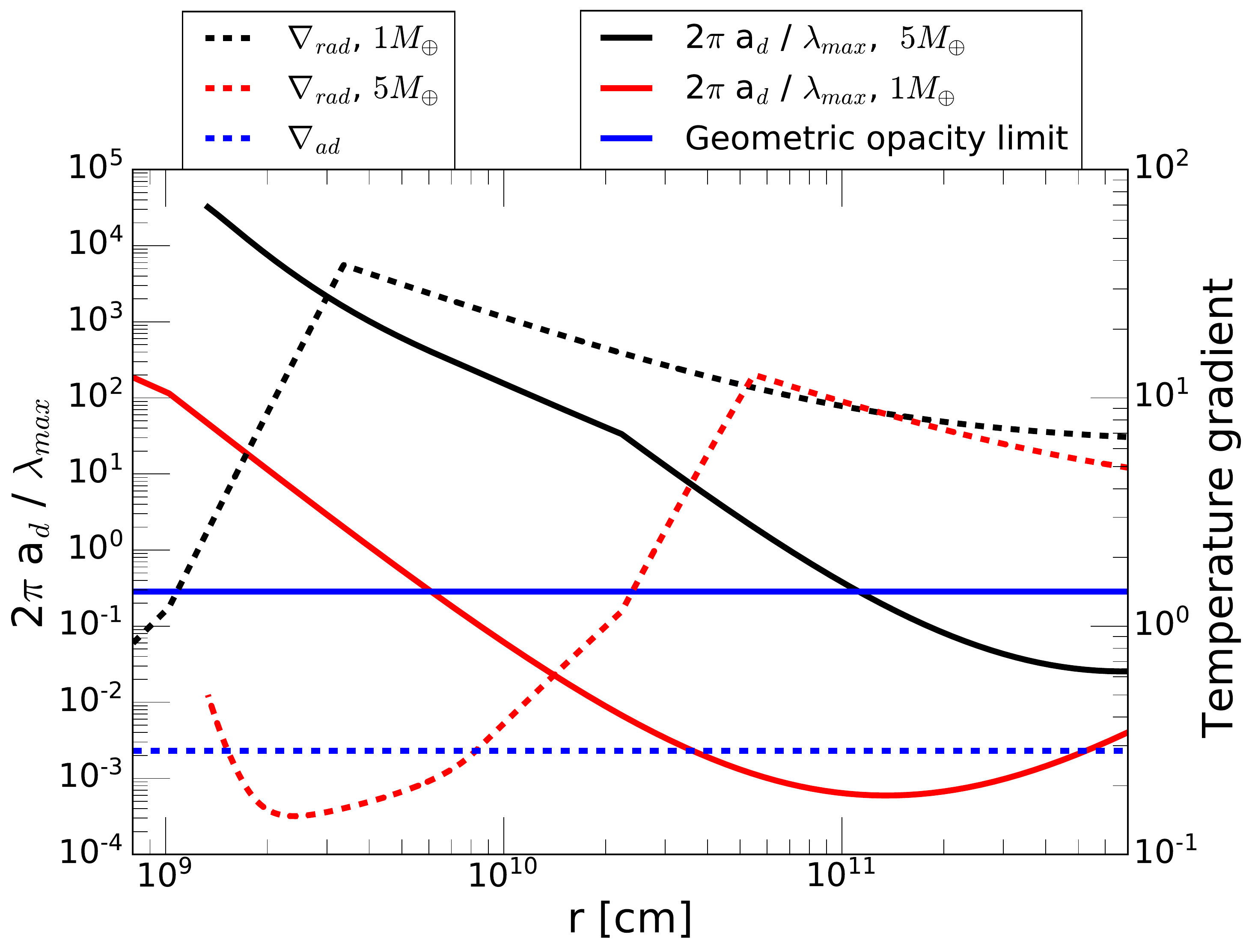}
	\includegraphics[scale=0.250]{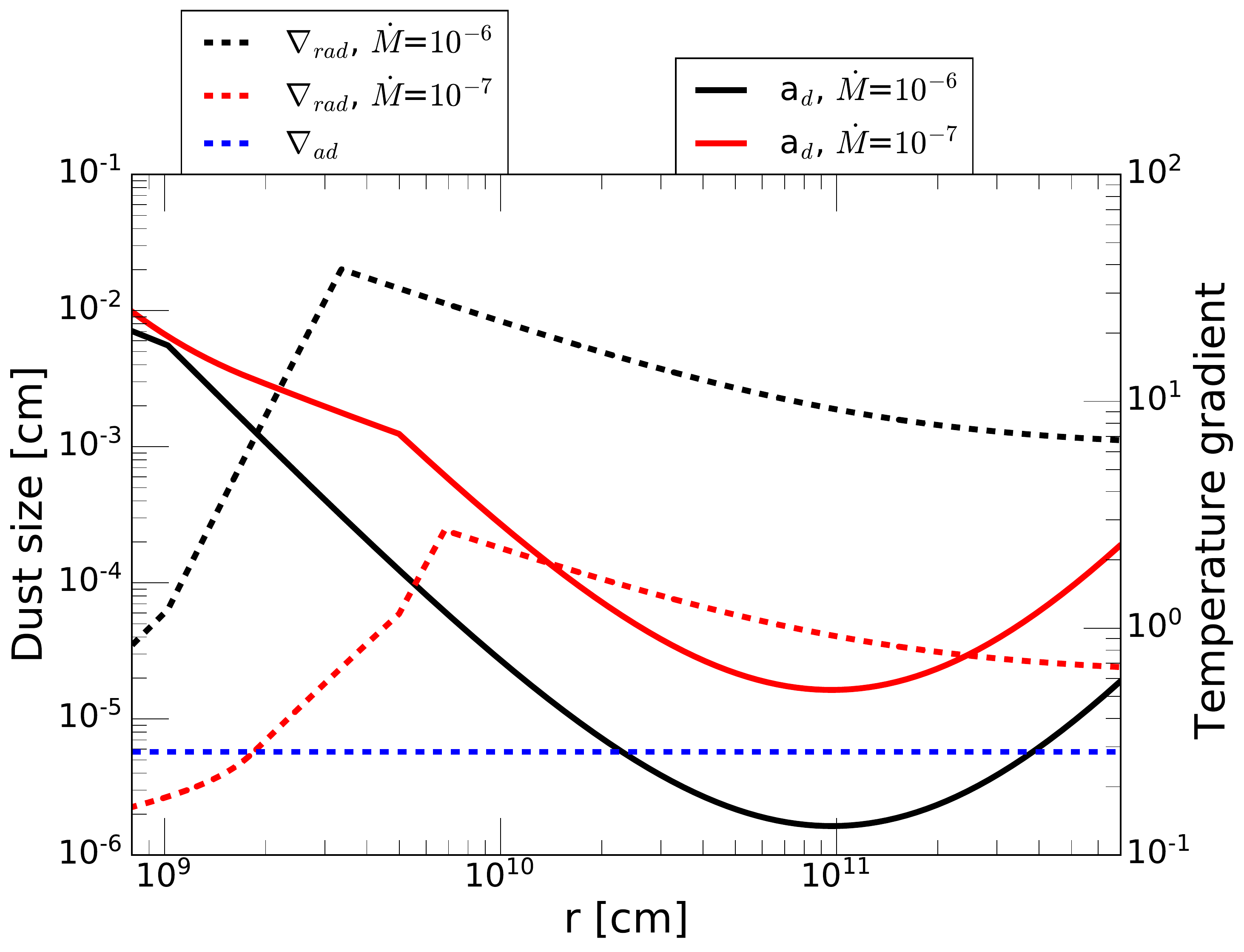}
   \caption{{Left:} solid lines are the opacity efficiency factors $Q_e$ (eq. \ref{grainopeq}) for 2 different core masses with all other parameters being equal {(15 AU, $\dot{M}_{acc} = 10^{-6} M_{\oplus}/yr$)}. These reach the regime switch value of 2 (solid blue line) at different radii, creating a radiative zone in the inner envelope for the 5 M$_\oplus$ case but not for 1 M$_\oplus$ due to its smaller dust size. The dashed lines are the radiative and adiabatic gradients, {indicating the radiative and convective zones}.   {Right:} solid lines are the dust size $a_d$ for cases with 2 different accretion rates but all other parameters being equal {(15 AU, 1 M$_\oplus$)}. Dashed lines are the radiative and adiabatic gradients for the same cases. {In all plots, the x-axis is the radius from the core, extending from the core to the envelope's outer boundary.}}
    \label{fig:main2}
    \end{centering}
\end{figure*}

Figure \ref{fig:main} shows that, while CSs can be created under a variety of parameter ranges, trends do exist. We start with our nominal model, for $\dot{M}_{acc} = 10^{-6} M_{\oplus}/yr$.  First, there is a relatively narrow range of masses that extends from around 0.08 M$_\oplus$ (40 times Pluto's mass) to 1.5 M$_\oplus$, beyond which the chances of creating CSs drops drastically.  This implies that  CSs might have formed in the proto-envelopes of Mars to Earth mass protoplanets that have since disappeared via giant collisions or dynamical ejection, or possibly grown into giant planetary cores. The lower limit on core masses is mainly due to their envelopes' relatively cooler temperatures, increasing $\tau_{vap,sil}$ considerably. On the other hand, cores with masses higher than 1.5 M$_\oplus$ have dust particles large enough in their middle and inner envelopes to switch from the Rosseland mean opacity regime to the geometric opacity regime, as can be seen in Fig. \ref{fig:main2} (left hand panel). This decreases the radiative gradient, creating an inner radiative zone that prevents these envelopes from being fully convective. {It is worth noting that, while our model considers the smallest of the Hill and Bondi radii to be the envelope's outer boundary, all of our acceptable cases that form CSs are in the Bondi regime. This is expected as the Hill regime dominates for higher mass cores (5-10 M$_\oplus$) that were excluded above. The Bondi radius $R_B = 2GM/c_s^2$ is obtained by equating the local sound speed to the gravitational escape velocity, and thus describes a usually light but bound envelope where gas particles do not have enough thermal energy to escape. For higher mass cores, the Bondi radius is large enough for the Hill stability criteria to become the more stringent constrain. }

\subsection{Semimajor axis}

A complementary piece of information is the semimajor axis, where we find that CSs can form almost anywhere in the disk if the envelope's grain abundance is high enough as discussed further below. Semimajor axis controls the temperature and density at the outer boundary, which seems to be important only in the marginal cases, for example for low core masses where the envelopes would be too cold if placed further out in the disk. 
The wide range of possible semimajor axis allows for the possibility of creating CSs in the comets formation region. Classically, Oort cloud comets were thought to form among the giant planets all the way down to 5 AU, while Jupiter family comets were thought to form in the scattered disk \citep{duncan,duncan2,dones}. Alternatively, \cite{brasser} proposed that both could have formed in the same region beyond Neptune. 

\subsection{Grains abundance}

{We moreover find that creating CSs necessitate solar to supersolar grain abundance in the envelope ($Z_{gr}$ >= 0.01). This result is not consistent with the subsolar grain abundances found in models that incorporate dust growth \& settling to the core but omit dust fragmentation with convective mixing. \cite{ormel} for example added a simple grain growth equation to the atmospheric structure equations, and found that $Z_{gr}$ can be as low as $10^{-4}$ in parts of the envelope. \cite{morda} also created an atmospheric model incorporating dust settling and coagulation, and found that this mostly results in subsolar opacities.} The main role of $Z_{gr}$ is to increase the opacity and extend the convective zone all the way to the outer boundary (Bondi or Hill radius). \cite{alidibthompson} discussed the gradual buildup of $Z_{gr}$ in the envelope through accretion and fragmentation, and derived a lower limit on $Z_{gr}$ in order to get a fully convective envelope:

\begin{equation}
Z_{gr}>0.12 \frac{T_{d, 2}^2}{\rho_{d,-11}}\left(\frac{t_{\mathrm{acc}, c}}{\mathrm{Myr}}\right)\left(\frac{M_c}{0.3 M_{\oplus}}\right)^{-2 / 3}
\end{equation}
which is generally consistent with our $Z_{gr}$ values. In order to get supersolar $Z_{gr}$, multiple conditions need to be satisfied:
\begin{itemize}
\item The dust size need to be fragmentation-limited, which is a pre-requisite for our dust-size prescription. This depends on many factors, including the accretion rate (setting the luminosity and thus convective speeds) and particles' porosity and chemical composition \citep{blum,Okuzumi,wada1,wada2}. 
\item A significant fraction of the dust should not get accreted by the core, but remain mixed in the envelope. This is an open question with many complications. In our cases, silicates are in vapor form at the base of the envelope which should stop accretion from taking place unless the temperature is low enough for the inner envelope to reach saturation pressure. This also depends on the nature of convection in these envelopes, whether it is diffusive as we are assuming, or whether it is dominated by large scale eddies that can enhance accretion by the core \citep{johansen}.
\end{itemize}

\subsection{Accretion rate}
Finally we investigate the effects of using a lower accretion rate. Our results for $\dot{M}_{acc} = 10^{-7} M_{\oplus}/yr$ are shown in Fig. \ref{fig:main}. In this case we find that while the semimajor axis range remains the same and the lower mass limit does not change ($\sim 0.08 M_{\oplus}$), the upper limit decreases by over a factor 2 to $\sim 0.6 M_{\oplus}$. This is expected since, lower accretion rate leads to lower luminosities. As seen in Fig. \ref{fig:main2} (right hand side), this decreases the radiative gradient and allows for a radiative zone in the inner envelope even though the dust size in the 2 cases converge to the same inner value. In some cases $Z_{gr}$ can compensate for the lower luminosity and increases the opacity enough to create fully convective envelopes, explaining the overall larger $Z_{gr}$ we find for the lower accretion rate cases.

\section{Summary \& conclusions}
Crystalline silicates are ubiquitous in comets, but can only form at very high temperatures.
Here we investigated the possibility of transforming amorphous silicates into crystalline particles inside the envelopes of protoplanets through vaporization followed by re-condensation, and then ejecting them back to the disk through diffusion in the fully convective envelopes. Using a simplified 1D envelope structure model that incorporates a dust size prescription accounting for fragmentation and growth, we showed that crystalline silicates can be created from a diverse set of parameters. Cores need to be between 0.08 to 1.5 M$_\oplus$ in mass, as lighter cores do not allow for temperatures high enough to vaporize silicates, and the envelopes of more massive cores are often not fully convective. We finally found that the location in the disk (1 to 30 AU) has little influence on the results, except in marginal cases, and that a solar to supersolar grain abundance is needed, but this can be achieved through dust fragmentation and accumulation. Our mechanism is simple and does not rely on assumptions about the disk, although it depends on the assumed diffusive nature of 1D convection. Whether this is realistic needs to be investigated further using 3D hydrodynamic simulations.

\section*{Acknowledgements}

We thank the anonymous referee for their constructive comments that greatly improved this manuscript. This work is supported by Tamkeen under the NYU Abu Dhabi Research Institute grant CAP$^3$.

\section*{Data availability}
The data underlying this article (numerical simulations output files) will be shared on reasonable request to the corresponding author.








\appendix


\bsp	
\label{lastpage}
\end{document}